\RequirePackage{fix-cm}
\documentclass[smallextended]{svjour3}       
\smartqed  
\usepackage{graphicx,algorithm, algorithmic}

\usepackage[numbers,sort]{natbib}
\usepackage{color}
\graphicspath{ {images/} }
\begin{document}

\title{Discovering the hidden community structure of public transportation networks
}


\author{L\'aszl\'o Hajdu \and
       Andr\'as B\'ota         \and
       Mikl\'os Kr\'esz \and
       Alireza Khani         \and
       Lauren M. Gardner
}


\institute{L. Hajdu \at
	     University of Szeged \\
	    Faculty of Science and Informatics \\
	    P. O. Box 652., \\
	    6701 Szeged, Hungary\\
               \email{hajdul@inf.u-szeged.hu}
	\and
	   A. B\'ota \at
	   Research Centre for Integrated Transport Innovation \\
	   School of Civil and Environmental Engineering \\
	   University of New South Wales \\
	   Sydney NSW 2052, Australia \\
		{\em also at} \\
           Integrated Science Lab \\
	   Department of Physics  \\
           Ume\r{a} University \\
           SE-901 87 Ume\r{a}, Sweden \\
              \email{andras.bota@umu.se} \\
	\and
	 M. Kr\'esz \at
	 University of Szeged \\
	 Gyula Juh\'asz Faculty of Education \\
	 Boldogasszony sgt. 6 \\
	 6720 Szeged, Hungary\\
               \email{kresz@jgypk.szte.hu} \\
		 {\em also at} \\
	 Innorenew CoE \\
	 Livade 6, \\
	 6310 Izola, Slovenia\\
     \email{miklos.kresz@innorenew.eu}     
           \and
           A. Khani \at
           Department of Civil, Environmental and Geo- Engineering \\
           University of Minnesota Twin Cities \\
           500 Pillsbury Drive SE, Minneapolis, MN 55455, USA \\
           \email{akhani@umn.edu}
	\and
            L. M. Gardner \at
	   Research Centre for Integrated Transport Innovation \\
              School of Civil and Environmental Engineering \\
	   University of New South Wales \\
	   Sydney NSW 2052, Australia \\
	  	\email{l.gardner@unsw.edu.hu} \\
		{\em also at} \\
            Department of Civil Engineering \\
            Johns Hopkins University \\
            Baltimore, MD 21218, USA \\
         \and
}

\date{Received: date / Accepted: date}

\maketitle

\begin{abstract}

Advances in public transit modeling and smart card technologies can reveal detailed contact patterns of passengers. A natural way to represent such contact patterns is in the form of networks. In this paper we utilize known contact patterns from a public transit assignment model in a major metropolitan city, and propose the development of two novel network structures, each of which elucidate certain aspects of passenger travel behavior. We first propose the development of a transfer network, which can reveal passenger groups that travel together on a given day. Second, we propose the development of a community network, which is derived from the transfer network, and captures the similarity of travel patterns among passengers. We then explore the application of each of these network structures to identify the most frequently used travel paths, i.e., routes and transfers, in the public transit system, and model epidemic spreading risk among passengers of a public transit network, respectively. In the latter our conclusions reinforce previous observations, that routes
crossing or connecting to the city center in the morning and afternoon peak hours are the most ``dangerous'' during an outbreak.

\keywords{network modeling \and public transportation \and community structure \and infrastructure security}
\end{abstract}

\section{Introduction and background}

Networks can be used to represent public transportation systems from various unique perspectives. Traditionally nodes and edges represent the physical infrastructure of a transportation system, e.g., routes and stops/stations \cite{london}, while recent developments in data collection and modeling allow researchers to accurately map contacts between individuals traveling together. Detailed commuting patterns can be recorded from smart card data \cite{sun,sun2} or can be the output of activity-based travel models \cite{l32,l33,l34,l35,l36}. The resulting passenger contact patterns are both spatial and temporal in nature, and include travel times on specific vehicles and contact with other travelers \cite{l41,l42,l64}. Such detailed travel patterns can then be used for various planning objectives including the estimation of the capacity of infrastructure \cite{WANG2011814}, calculating environmental impact \cite{CARLSSONKANYAMA1999405} or designing surveillance and containment strategies during an epidemic outbreak \cite{l49}. While these methods  allow researchers to map contacts between known individuals, the data collection and processing required to recreate a real-world contact network presents many challenges in terms of accuracy and computational complexity, among other issues such as privacy \cite{l43, l44, l45, l46, l47, l48}.

In this work we specifically focus on the community structure of public transit ridership patterns.  Observations on social interactions reveal that people tend to form groups according to their lines of interest, occupation, etc. This concept is known as homophily in the social sciences \cite{eagle2009inferring,yuan2006homophily,chin2012using}, while in network modeling we refer to this phenomenon as community structure \cite{fortunato}. One of the most frequently used definitions of this concept was proposed by Newman \cite{newman}: ``In an arbitrary network a {\em community} is a set of nodes where the density of connections between the nodes of the community is greater than the density of connections between communities''. Community detection is a diverse field with many applications in various fields of science. Newman's definition has been extended or replaced by newly proposed algorithms, yet it is still the most intuitive way of describing communities.

The majority of community detection algorithms consider communities as disjoint sets of nodes. Popular approaches include modularity maximization methods \cite{newman,blondel}, information theoretic approaches \cite{infomap}, statistical inference \cite{sbm,wsbm} and spectral techniques \cite{krzakala,newmanspectral}.
Other algorithms allow overlaps between neighboring groups \cite{bota2015high,kertesz,wu}. Community detection algorithms can also be applied to weighted networks, where a value on each edge represents similarity (or distance) between nodes \cite{bota2015high,wsbm}. All weighted networks can be transformed into unweighted forms by setting a threshold and omitting edges with weights below the threshold. This technique has been used in real-world applications \cite{palla2005uncovering,bota2014community}. A selection of excellent reviews on community detection can be found in \cite{fortunato,overlapping,leskovec}.

In this work we propose the development of two novel network structures, namely the {\em transfer network}, and the {\em community network}. The transfer network captures the movement patterns of atomic passenger groups, i.e., groups of people who travel together on one or more vehicles, mathematically defined as maximal complete subgraphs of the contact network. The community network captures the
similarity of travel patterns between passengers, is derived directly from the transfer network, and is constructed using a novel link-based metric called the {\em connection strength}. The community network is intended to reveal groups of travelers who may also be more likely to come into contact outside of their daily travel routine e.g.: colleagues traveling to work or children going to school \cite{yuan2006homophily,chin2012using}. A description of each proposed network structure as well as the contact network can be found in Table 1.
	
We further demonstrate potential applications of each of these novel network structures. First, the transfer network is implemented to detect the most frequent vehicle trip combinations in the system (bus and train transfers). While there are existing methods in the literature to measure the capacity of public transit systems \cite{cepeda2006frequency,lai2011behavioral}, our approach provides a more refined view of passenger movement between vehicle trips by tracking the movements of groups of passengers traveling together. Identifying the most relevant vehicle trip combinations can aid public transit authorities in timetable planning and optimizing vehicle assignments. 

Second, the community network is applied to evaluate a diffusion process in a transit network, specifically infectious disease spread among passengers. This application complements our previous work \cite{nets}, to identify the components of the public transportation system most vulnerable to a bio-security threat. The community network proposed here can further improve the performance of such models due to its novel representation of passenger interactions. We use the proposed network structure to simulate how a disease might spread among the vehicles of the public transportation system and the suburbs of the city, and identify the vehicle trips most likely to carry infected passengers. All applications are evaluated on a real-world case study, the public transit system of Twin Cities, MN, using output from an activity based travel demand model \cite{l64}.

The rest of the paper is structured as follows. In Section 2 we introduce the travel demand model, and how it is used to create passenger contact network. In Section 3 we introduce the transfer network, and illustrate how it can be used to detect frequent vehicle trip combinations. In Section 4 we introduce the community network, including the connection strength value, and illustrate how it can be used to model infectious disease spread. Section 5 contains our conclusions and future research directions.

\begin{table}[!h]
	\begin{center}
		\caption{Network definitions used in this paper, including the contact network and the proposed transfer and community networks}\label{tab:test}
		\begin{tabular}{|p{1,2cm}|p{3cm}|p{3cm}|p{3cm}|} \hline
			 & Contact network & Transfer network & Community network  \\ \hline
			Nodes & Passengers &  Atomic passenger groups &  Passengers traveling on at least two vehicle trips \\ \hline
			Edges & Passengers are connected of they are physically present on the same vehicle trip at the same time (undirected) & Atomic passenger groups are connected if they share a passenger (directed)
 & Passengers are connected if they are traveling together on at least two vehicle trips (undirected) \\ \hline
			Attributes & Contact duration and start time & Number of transfer passengers between atomic groups & Connection strength measuring the similarity of the travel patterns between passengers  \\ \hline
		\end{tabular}
	\end{center}
\end{table}

\section{Model Inputs}

The inputs of our work are based on the transit assignment model published in \cite{l64}.
In this section, we present a description of the assignment model and provide a short summary of the properties of the passenger contact network built from it.

\subsection{Travel demand model}

As described in our previous work \cite{nets}, public transportation data in this study was obtained from the transit system in Twin Cities region in Minnesota, where 187 routes serve 13,700 stops in the region. Transit network and schedule data were created from General Transit Feed Specification (GTFS), including near 0.5 Million stop-times on a weekday in 2015. Transit passenger trips were obtained from Metropolitan Councils's activity-based demand model \cite{alireza}, and contained more than 293,000 linked trips (i.e. a passenger trip from an origin to a destination that may include zero or more transfers). The assignment of transit demand to transit network was done using the FAST-TrIPs model \cite{l64}. FAST-TrIPs is a schedule-based transit assignment model that generates hyperpaths using a defined logit route choice model \cite{alireza2}, assigns individual passengers to the paths using the hyperpath probabilities, and simulates them using a mesoscopic transit passenger simulation module. Since a calibrated transit route choice model was not available for the Twin Cities network, a route choice model from Austin, TX was borrowed from a previous study by the authors \cite{alireza3}. The model specifies the following route choice utility function:
$$u = t_{IV} + 1.77t_{WT} + 3.93t_{WK} + 47.73X_{TR}$$
where $t, t_{IV}, t_{WT}, t_{WK}$ and $X_{TR}$ represent path utility, in-vehicle time, waiting time, walking time, and number of transfers in a transit path, respectively. The output of the transit assignment model contains individual passengers' trajectories, including their walking from the origin to the transit stop, boarding the transit vehicle, alighting and walking to the transfer stop, boarding the next transit vehicle, etc. and finally alighting and walking to the destination. By post-processing the transit assignment model's outputs, the amount of time each pair of passengers are on-board the same transit vehicle were calculated on a daily basis.

\subsection{Contact network}

We define a network structure denoted as the {\em contact network} based on the outputs of the travel demand model. The nodes of the contact network are passengers, and edges connect passengers if they were traveling on the same vehicle at the same time. The relationship is symmetric, therefore the network is undirected. All passenger movements take place in a temporal setting, which is indicated by two values assigned to the edges of the network: the contact start time on an edge indicates the start time of the contact between the two connected passengers, while a contact duration value indicates the length of the contact in minutes. Since each edge represents a connection between a pair of passengers traveling on a specific vehicle, the id of the vehicle can also be assigned to the edges of the graph. The vehicle id identifies a vehicle trip: a single vehicle traveling on a specific route with a specific start time. The basic properties of the network were shown in \cite{nets} along with a detailed analysis regarding the networks potential role in the spreading of epidemics. A short review on the description of the network as appeared in \cite{nets} follows.

\begin{figure}[tbp]
	\centering
	\includegraphics[scale=0.34]{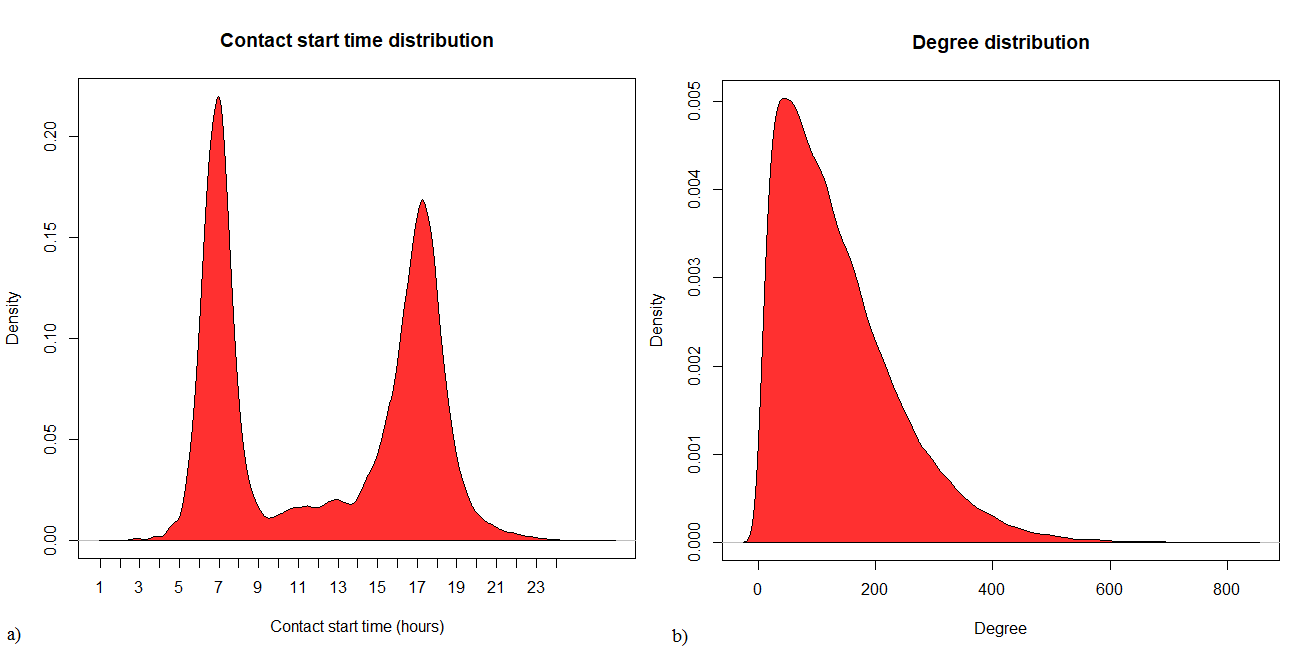}
	\caption{The distribution of a) contact start times and b) the degree distribution of the contact network.}
\end{figure}

\begin{figure}[tbp]
	\centering
	\includegraphics[scale=0.34]{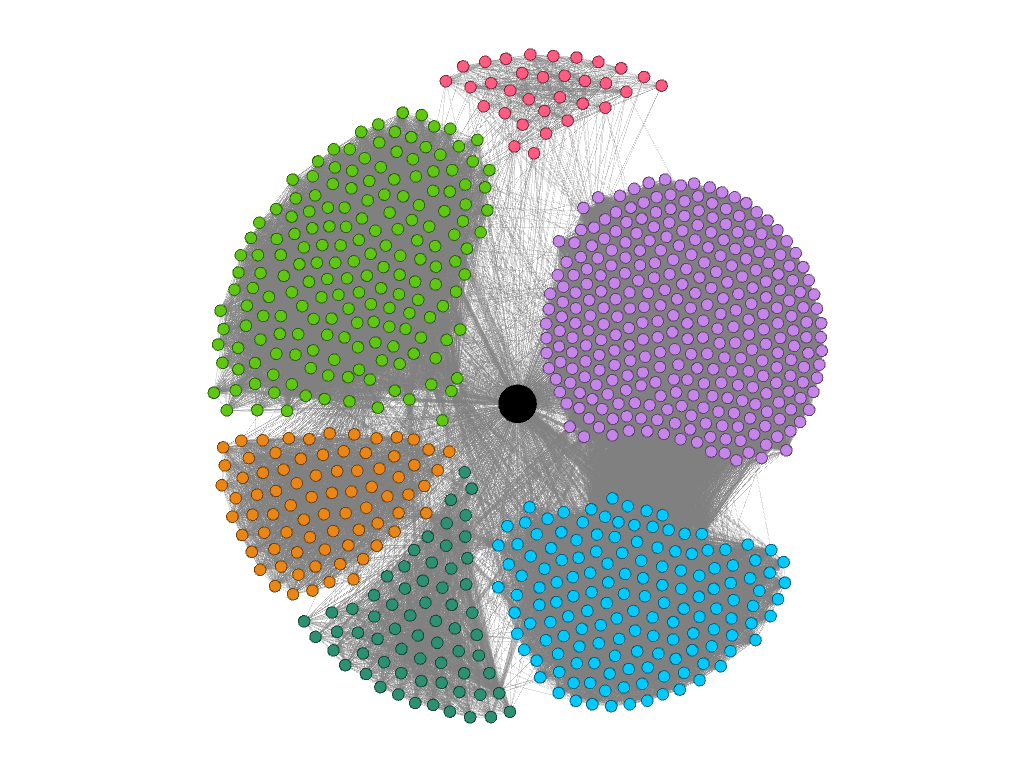}
	\caption{A static subgraph of the contact network. A passenger with a high number contacts is represented by the black node in the middle, connected to all other contacts. The colors indicate of the vehicle trips the passengers first met on. The figure was made with Gephi using the Fruchterman-Reingold algorithm.}
\end{figure}

The contact network corresponding to the Twin cities dataset has $94475$ nodes and $6287847$ contacts between them. Figure 1/a shows the density plot of the contact start times. The  distribution has two peaks, one around 7 AM and another around 5 PM, which corresponds to the morning and evening weekday commute. The average number of contacts per person is 136 during the observed day, while the maximum is 827. Figure 1/b shows the degree distribution of the graph.

Figure 2 shows a subgraph of the network structure. Nodes represent passengers and edges connect them if they ride on the same vehicle trip. The dynamics of the network are omitted in this example, i.e. edges are aggregated over the entire day. The black node in the middle represents a passenger with a high number of contacts, who traveled together with all the other nodes shown on this subgraph. The other nodes are colored according to the vehicle trip they rode on, while darker edges indicate longer contact durations.

\section{Transfer network}

The first objective of this paper is to detect the movement patterns of passenger groups.
To do this, we define a  novel network structure denoted as the {\em transfer network}. The transfer network is a directed network, where nodes represent the atomic passenger groups of the contact network and groups are connected if at least one member of a group transfers from one vehicle to another. The weight of the edges denote the number of transfer passengers.

In order to build the transfer network we identify the subgraph corresponding to each vehicle trip, detect atomic passenger groups -- defined as maximal cliques -- on each of the resulting subgraphs of the contact network and connect the atomic passengers groups according to direction of transfer between vehicle trips. We weight the connections based on the number of transferring passengers. Section 3.1 defines the construction of the transfer network in more detail.

The transfer network proposed in this paper provides a refined way to identify the most frequent vehicle trip combinations in the public transit system, which can aid decision makers in defining timetables and optimizing vehicle assignments. Section 3.2 discusses this application.

\subsection{Transfer network construction}

The three steps required to build the transfer network are discussed in the subsequent subsections. Section 3.1.1 defines how the subgraphs corresponding to the vehicle trips are constructed, 3.1.2 defines atomic passenger groups as cliques and show how they can be detected in an efficient way and 3.3.3 shows how the transfer network can be built from the passenger groups.

\subsubsection{Graph partitioning}

We define the subgraph corresponding to each vehicle trip as follows. Let $G$ be the original contact network, $V(G)$ the vertices and $E(G)$ the edges of the network. We divide the original network into subgraphs along vehicle trips. Let $T$ be the set of all vehicle trips in the network and let $t_i \in T$ denote the $i$-th trip.  We define $G_{t_i}$ as the subgraph corresponding to trip $t_i$ where, $V(G_{t_i})$ and $E(G_{t_i})$ are the vertices and the edges of trip $G_{t_i}$. We create a subgraph $G_{t_i}$ for all trips  $t_i \in T$. Since a single passenger may travel on multiple trips, the node corresponding to the passenger may appear in multiple subgraphs. Figure 3 shows an example of the partitioning process.

\begin{figure}[h]
	\centering
	\includegraphics[width=1\textwidth]{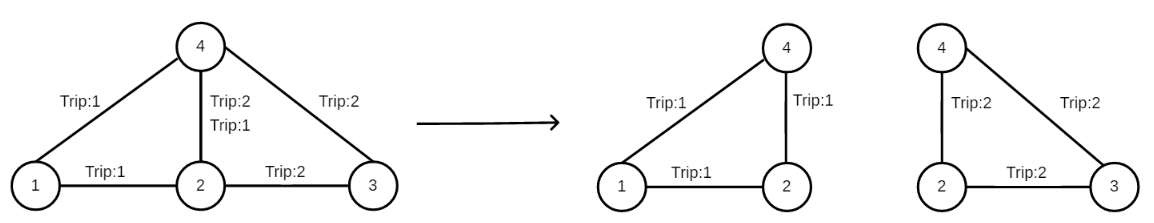}
	\caption{Partitioning of an example graph along vehicle trips. Each vehicle trip has a corresponding subgraph where nodes are passengers who used the given vehicle trip.}
\end{figure}

\subsubsection{Clique detection}

In graph theory, a {\em clique} is defined as a fully connected subgraph of a given graph. A {\em maximal clique} is a clique, that is not a subgraph of any other clique. Finding the set of all maximal cliques is a well-studied NP-hard problem in graph theory. An arbitrary $n$-vertex graph may have up to $3^{n/3}$ maximal cliques, but this number is much lower in many complex networks, including the contact network studied in this paper. Among the available methods in clique detection we adopt the Bron-Kerbosch (BK) algorithm \cite{Bron:1973:AFC:362342.362367}, which has proven its reliability in real-life applications \cite{eppstein2010listing}. We apply the BK algorithm\footnote{We implemented the Bron-Kerbosch algorithm with pivoting and degeneracy ordering in the outer most level of recursion in C++. We ran the algorithm on a PC with I7 4790 CPU (3.6 Ghz) and 16GB of RAM. } to detect the maximal cliques of all subgraphs $G_{t_i}$ corresponding to all vehicle trips  $t_i \in T$. 

The analysis can be further extended if we consider the graphs as weighted according to the contact durations available on the edges. Introducing a weight threshold $\tau$ we can prune the edges of the graphs by omitting all edges with contact durations below $\tau$. This allows us to redefine what a connection means in the network: passengers are only considered to be connected if they spend at least a certain amount of time on the same vehicle. 
We define three thresholds to represent the strength of connection between passengers $\tau_5 = 5$ minutes, $\tau_{15} = 15$ minutes and $\tau_{30} = 30$ minutes. We chose these thresholds to represent short, medium and long duration contacts between passengers.
We prune the edges of the contact network by these values resulting in graphs $G_5$, $G_{15}$ and $G_{30}$. In additional, let $G_0$ and $\tau_0$ represent the original (uncut) graphs and the corresponding threshold. We run the Bron-Kerbosch clique detection algorithm on these graphs and analyze and compare results. The speed of the detection algorithm is amplified by the fact, that all subgraphs corresponding to vehicle trips are interval graphs. In order to show this speedup, we also run the clique detection algorithm on the original contact network and compare results. Table 2 shows graph size, runtime of the clique detection method on the original contact network and the total runtime of the detection method on all subgraphs for $G_0$, $G_5$, $G_{15}$ and $G_{30}$. We can see a significant speedup for all graphs in this analysis. For the larger $G_0$ and $G_5$ graphs the total runtime on all subgraphs corresponding to vehicle trips is 14 times less than on the unpartitioned network.

\begin{table}[!htb]
	\begin{center}
		\caption{ Runtimes of the Bron-Kerbosch on graphs $G_0, G_5, G_{15}, G_{30}$. Raw denotes runtime in seconds on the original contact network, while partitioned denotes the total runtime on all subgraphs corresponding to vehicle trips.}\label{tab:test}
		\begin{tabular}{r|c c c c c c} \hline
			\begin{tabular}[x]{@{}c@{}}Graph\end{tabular}& \begin{tabular}[x]{@{}c@{}}Passengers\end{tabular}&\begin{tabular}[x]{@{}c@{}}Edges\end{tabular}&\begin{tabular}[x]{@{}c@{}}Trips\end{tabular}&\begin{tabular}[x]{@{}c@{}}Time\\raw (s)\end{tabular}&\begin{tabular}[x]{@{}c@{}}Time\\partitioned (s)\end{tabular}\\ \hline
			$G_0$ & 94475   & 6435482 & 8002 & 5285 & 367   \\ 
			$G_5$ & 91894   & 5203557 &  7934 & 3830 & 258  \\ 
			$G_{15}$ & 63714   & 1630522 & 6969 & 1195 & 50  \\ 
			$G_{30}$ & 26154   & 218233 & 4083 & 129 & 6  \\ 
			
		\end{tabular}
	\end{center}
\end{table}

\subsubsection{Graph building}

Let $F$ denote the {\em transfer network} as a directed graph where every node $v \in V(F)$ is a clique from $G_{t_i}$ for all $t_i$. Edges connect nodes $v$ and $u$ if the corresponding cliques do not lie in the same vehicle trip, yet they have at least one common passenger. More formally, for nodes $u$ and $v$ in the transfer network and their corresponding cliques $c_v$ and $c_u$ in the contact network, $u$ and $v$ is connected if $c_v \cap c_u \geq 1$ and if $c_v \subseteq G_{t_v}, c_u \subseteq G_{t_u}$ then $G_{t_v} \neq G_{t_u}$. 

The {\em direction of edges} correspond to the direction of the transfer between $t_u$ and $t_v$, that is whether the passengers of $c_v \cap c_u$ move from $t_u$ to $t_v$ or the opposite. We establish the direction by looking at the contact start times for the individuals in $c_v \cap c_u$ in the following way. For all edges of the transfer network $e_{uv} \in E(F)$, let $c_{uv}$ denote the set of corresponding passengers in $G$ as $c_{uv}=c_u \cap c_v,  c_v \subseteq G_{t_v}, c_u \subseteq G_{t_u}$, $G_{t_v} \neq G_{t_u}$. Let $\alpha_{xy}^{t_{i}}$ denote the contact start time between passengers $x$ and $y$ on vehicle trip $t_i$. For all passengers $p,q \in c_{uv},$ if $min(\alpha_{pq}^{t_{u}}) < min(\alpha_{pq}^{t_{v}}) $, then the direction of the edge $e_{uv} \in E(F)$ is from $u$ to $v$, else it is from $v$ to $u$. If $|c_v \cap c_u| = 1$, then let $c_v \cap c_u = \{p_0\}$ and $p \in c_u \setminus \{p_0\}, q \in c_v \setminus \{p_0\}$. Then $min_p(\alpha_{p_{0}p}^{t_{u}}) < min_q(\alpha_{p_{0}q}^{t_{v}})$ decides the direction of the edge as above.
We assign integer values $w_v = |c_v|$ to all $v \in V(F)$ and $w_{u,v} = |c_v \cap c_u|$ for all $e(u,v) \in E(F)$. Values $w_v$ and $w_{u,v}$ represent the amount of passengers corresponding to both the nodes and edges of the transfer network. 

\subsection{Detecting frequent vehicle trip combinations}

 The {\em transfer network} allows us to identify the most frequent vehicle trip combinations passengers travel on in the public transportation system. 
Detecting the most frequent combinations helps decision makers in defining timetables and optimizing vehicle assignments.

As before, let $T$ be the set of vehicle trips, and $t_i \in T$ the i-th vehicle trip. We define vehicle trip pairs as follows: for all $t_i$ and $t_j$, $t_{ij}$ is a vehicle trip pair where $i \neq j $ and $t_i,t_j\in T$, while the set of all vehicle trip pairs is denoted by $T^{p}$. We assign a number $m_{ij}$ to each vehicle trip pair $t_{ij} \in T^{p} $ indicating the amount of passenger traffic between the corresponding trips. To calculate this value let the set $V(C_{t_{ij}})$ contain all nodes of the transfer network corresponding to all cliques $c_{t_{ij}} \in C_{t_{ij}}$ where $c_{t_{ij}} \in G_{t_i} \cup G_{t_j}$, and take the subgraph induced by $V(C_{t_{ij}})$. The number $m_{ij}$ is the sum of all edge weights of the induced subgraph. This approach offers a more refined view of passenger traffic between vehicle trips because it represents the movements of passenger groups as opposed the behavior of individual passengers.  
 

To give another dimension to our analysis we compared the frequent vehicle trip combinations in both $G_0, G_5, G_{15}$ and $ G_{30}$, that is only considering passengers to be in contact with each other if they travel together for more than $\tau_5 = 5$ minutes, $\tau_{15} = 15$ minutes and $\tau_{30} = 30$ minutes in addition to the unfiltered network $G_0$. Table 3 shows the five most frequent vehicle trip combinations $t_{ij}$ and the amount $m_{ij}$ of passenger traffic between them.

\begin{table}[!h]
	\begin{center}
		\caption{The five most frequent vehicle trip combinations in $G_0, G_5, G_{15}$ and $G_{30}$. The first number indicates the route number followed by the start time of the sepcific vehicle trip.}\label{tab:test}
		\begin{tabular}{r|c c c|r|c c c} \hline
			Graph & $t_i$ & $t_j$ & $m_{ij}$ & Graph & $t_i$ & $t_j$ & $m_{ij}$ \\ \hline
			$G_0 \ 1.$ & 614/5:25   & 675/5:30 &  38 & $G_{15} \ 1.$ & 68/6:19    & 94/7:04 & 21\\ 
			$2.$ & 68/5:46   & 94/6:25 &  35 & $2.$ & 94/18:32 & 68/18:52 & 19 \\
			$3.$ & 68/6:19   & 94/7:04 &  27 & $3.$ & 68/5:46 & 94/6:24 & 18 \\
			$4.$ & 901/6:18  & 415/7:02 &  25 & $4.$ & 614/5:16 & 675/5:35 & 16  \\
			$5.$ & 415/17:12  & 901/17:33 &  23 & $5.$ & 294/5:38  & 94/6:44 & 15  \\ \hline
			$G_5 \ 1.$ & 614/5:25  & 675/5:30 & 37 & $G_{30} \ 1.$ & 19/17:22 & 850/18:03 &  8  \\
			$2.$ & 68/5:46  & 94/6:25 & 35 & $2.$ & 54/18:02 & 68/18:23 &  4 \\
			$3.$ & 68/6:19  & 94/7:04 & 27 & $3.$ & 71/6:07 & 61/6:38 &  4  \\
			$4.$ & 901/6:18 & 415/7:02 & 25 & $4.$ & 5/0:30 & 901/2:07 &  4   \\
			$5.$ & 415/17:12  & 901/17:33  & 23 & $5.$ & 902/19:25 & 68/19:52 &  4 \\ \hline
		\end{tabular}
	\end{center}
\end{table}

Almost all trip combinations in Table 3 follow a similar pattern. Results for $G_0$ and $G_5$ are nearly identical. The most frequent combinations for $G_{15}$ connect two additional suburbs (Oakdale and Stillwater) to $G_0$ and $G_5$, but otherwise are identical. The vehicle trips pairs with the greatest amount of passenger traffic between them are mostly in the morning or afternoon peak hours. Almost all of the trip combinations link one of the outlying suburbs to the city center, indicating the daily commuting patterns of workers or students. Just the urban area of Twin Cities covers 2646 $km^2$ with a population of more than three million. This means that commuters trying to reach the city center from one of the outlying suburbs must travel on two or sometimes three separate routes to get to their destination. The pattern -- also shown on Figure 4 --  is the following. Commuters start the journey from one of the {\em smaller outlying suburbs} like Ridgedale (route 614), Inver Grove and other suburbs south of St. Paul (route 68) or Oakdale and Stillwater (route 294). Then they change vehicles in the transport hub of one of the {\em major outlying suburbs} (St. Paul, Minnetonka, Mall of America in Bloomington), and take an express service to the {\em city center} (routes 94, 675, 850, etc..).

While the frequent combinations for $G_{30}$ are similar, there are differences. While the suburbs these routes connect are different, we can see almost the same patterns as before: travel from one of the smaller outlying suburb to a major one and then to the city centre (combinations 71--61 and 5--901). One specific route (850) is one of the longest express bus route in the city and it includes a long section where the bus doesn't stop at all. One exception to the pattern is route 54, which connects St. Paul International airport to St. Paul and route 68 connecting to south St. Paul. In terms of time in addition to peak hours, we see late night services as well.

\begin{figure}[h]
	\centering
	\includegraphics[width=1.0\textwidth]{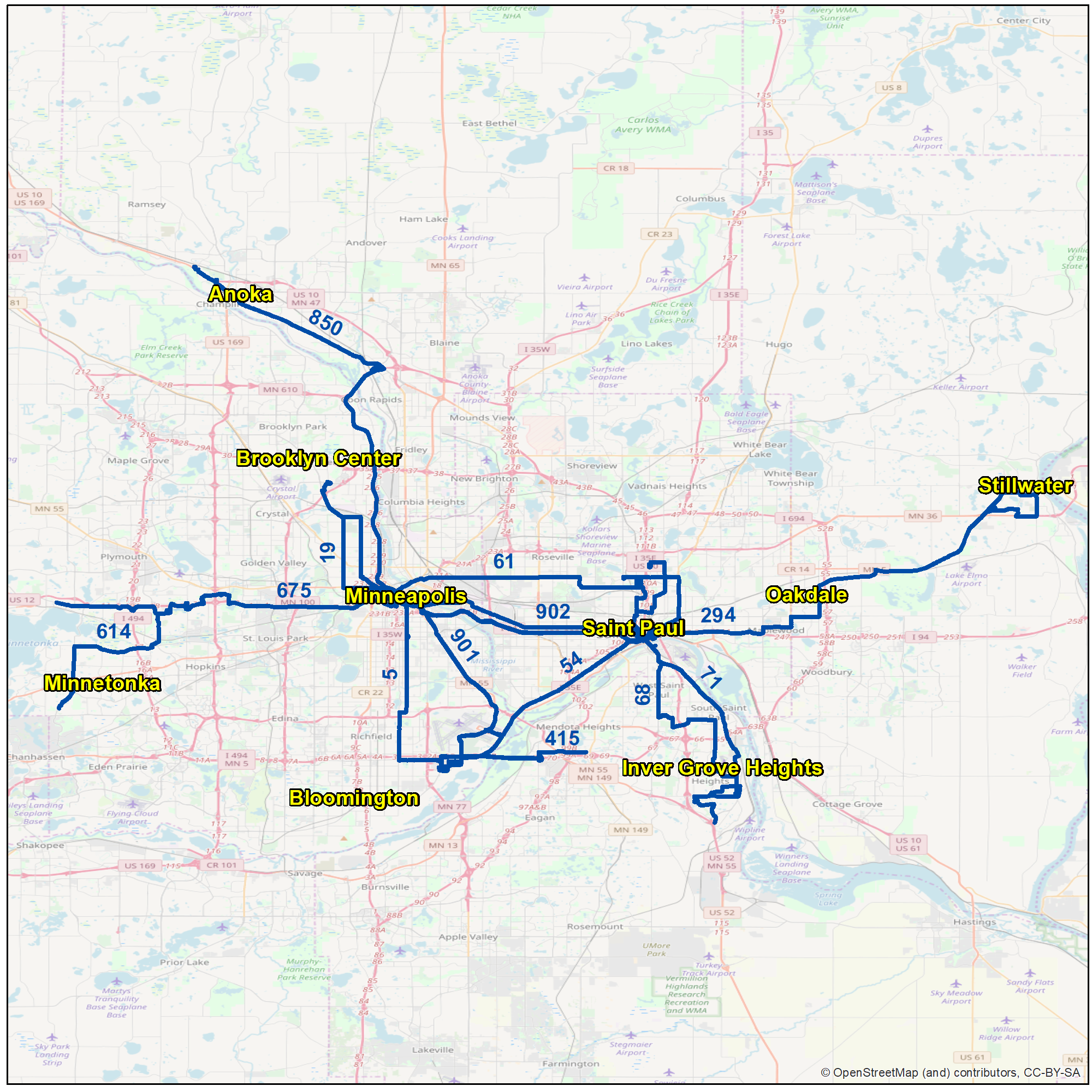}
	\caption{Most frequent vehicle trips combinations in Twin Cities, MN for $G_0, G_5, G_{15}$ and $ G_{30}$. }
\end{figure}

We can summarize, that graphs $G_0, G_5$ and $G_{15}$ behave similarly, showing the movements of passenger groups traveling through the public transit system. The most frequent trip combinations are the ones connecting distant suburbs to the city center during the morning and afternoon peak hours. $G_{30}$ captures an alternative set of routes corresponding to people who travel together for longer periods for different reasons, like a long service (route 850), the scarcity of services late at night (route 5 at 0:30) or traveling from the airport to a major transport hub (route 54).


\section{Community network}

Next we propose a novel network structure, the {\em community network}, which expands the definition of passenger connectivity to be a function of both the number of transfers passengers make together as well as the total amount of time they spend together while traveling. This contrasts the contact network, which simply quantifies passenger connectivity passengers using contact duration on individual vehicle trips. We define a {\em community of passengers} as a set of passengers who have common travel patterns, e.g., vehicle trips and/or transfers. 
In order to build communities, we define and quantify a  novel {\em connection strength} metric between passengers indicating the similarity of their travel patterns and create a new network structure, the {\em community network} based on this value. Section 4.1 outlines this process in more detail.

This network can serve as the basis of a community detection algorithm, but in this paper we take a different approach. We define communities as those connected by edges whose values lie above a predetermined threshold. In section 4.2 we demonstrate this feature by identifying the commuting patterns of the members of the largest passenger community of the network.

We propose an application of the community network in section 4.3. We show how the connection strength value can be used to model infectious disease transmission among passengers of a public transit system. Tying to our previous work in \cite{nets} we seek to identify the vehicle trips most likely to carry infected passengers during an outbreak.
 
\subsection{Community network construction}

In order to detect the communities of the public transportation network we construct a weighted network structure called {\em community network}.
The community network connects passengers using on a novel link-based metric, the {\em connection strength}. 
The connection strength $s$ defines the edge weights in the community network, and takes into account the number of transfers a pair of passengers makes together. Thus, if the passengers meet on multiple different vehicle trips, their connection strength $s$ will increase. This method is based on the assumption that passengers who not only travel together but also transfer together have a stronger connection than travelers who are simply present on the same vehicle trip at the same time. Below we explain how this link metric is derived.

Let $H$ denote the new network where the nodes are the passengers, and the set of nodes V(H) includes all passengers who traveled on at least two vehicle trips with another passenger. Thus, the nodes in the community network correspond to the edges of the transfer network. We define the connections and weights between passengers as follows. Nodes $u$ and $v$ in the community network are connected if they are both present in at least two different cliques $c_1, c_2$ in two different subgraphs corresponding to vehicle trips, \textit{i.e.} they traveled together on at least two vehicle-trips ($u,v \in c_1 \subseteq G_{t_1}$ and $u,v \in c_2 \subseteq G_{t_2}$). Let $g_{uv}$ denote the number of instances where $u$ and $v$ are members of the same clique, that is $g_{uv} = |C_{uv}|$ where $c_{uv} \in C_{uv}$ if $u,v \in c_{uv}$.
  Let $T_{uv}$ be the set of {\em vehicle trips} where both $u$ and $v$ are present: $t_{uv} \in T_{uv}$ if $u,v \in t_{uv}$, and let $g_{uv}^{t_{i}}$ be the number of the cliques in vehicle trip $t_{i}$ where $u$ and $v$ are both present: $g_{uv}^{t_{i}} = |C_{uv}^{t_{i}}|$ where $c_{uv}^{t_{i}} \in C_{uv}^{t_{i}}$ if $u,v \in c_{uv}^{t_{i}} \in G_{t_i}$. Thus, the connection strength $s_{uv}$ between passengers $u$ and $v$ can be formalized as follows:

\begin{equation}
s_{uv}=\frac{g_{uv}*(g_{uv}-1)}{2}-\sum_{t_{i} \ in \ T_{uv}}\frac{g_{uv}^{t_{i}}*(g_{uv}^{t_{i}} -1)}{2}
\end{equation}

Using this definition of connection strength, if a pair of passengers traveled together in $g_{uv}$ different atomic passenger groups, then they would have $\frac{g_{uv}*(g_{uv}-1)}{2}$ different edges between them because the affected nodes form a clique in the transfer network. This way the first part of the equation rewards the movements between different vehicle trips. Since based on our community definition traveling on the same vehicle trip doesn't indicate strong connection between the passengers, in the second part of the equation we penalize any instance where $u$ and $v$ travels together on the same vehicle trip. The value of the penalty will be the sum of the edges in every vehicle trip where the passenger pair appears more than one time, and the number of the edges in a vehicle trip is counted in the same way as in the first part of the equation. 


\begin{algorithm}[h]
	\caption{ Construction of the {\em community network}}  
	\renewcommand{\algorithmiccomment}[1]{\hfill {\it #1}}
	\begin{algorithmic}[1]
	\STATE $V(H) \leftarrow Every \ passengers \ from \ E(F) \ sections $ \
	\STATE $E(H) \leftarrow  \emptyset$ 
	\STATE \textbf{For} in $E(F) sections$ \textbf{Do}
	\STATE\hspace{\algorithmicindent} \textbf{For} $u \ v $ in every passenger pairs \textbf{Do}
	\STATE\hspace{\algorithmicindent}\hspace{\algorithmicindent} \textbf{If} {$e(u,v) \notin E(H) $}
	\STATE\hspace{\algorithmicindent}\hspace{\algorithmicindent}\hspace{\algorithmicindent} $E(H) \leftarrow e(u,v)$
	\STATE\hspace{\algorithmicindent}\hspace{\algorithmicindent} \textbf{Else}
	\STATE\hspace{\algorithmicindent}\hspace{\algorithmicindent}\hspace{\algorithmicindent} $s_{uv}++$
	\STATE\hspace{\algorithmicindent}\hspace{\algorithmicindent} \textbf{End if}
	\STATE\hspace{\algorithmicindent} \textbf{End for}
	\STATE \textbf{End for}

	\end{algorithmic}
\end{algorithm}

 \begin{figure}[h]
 	\centering
	\includegraphics[width=1.0\textwidth]{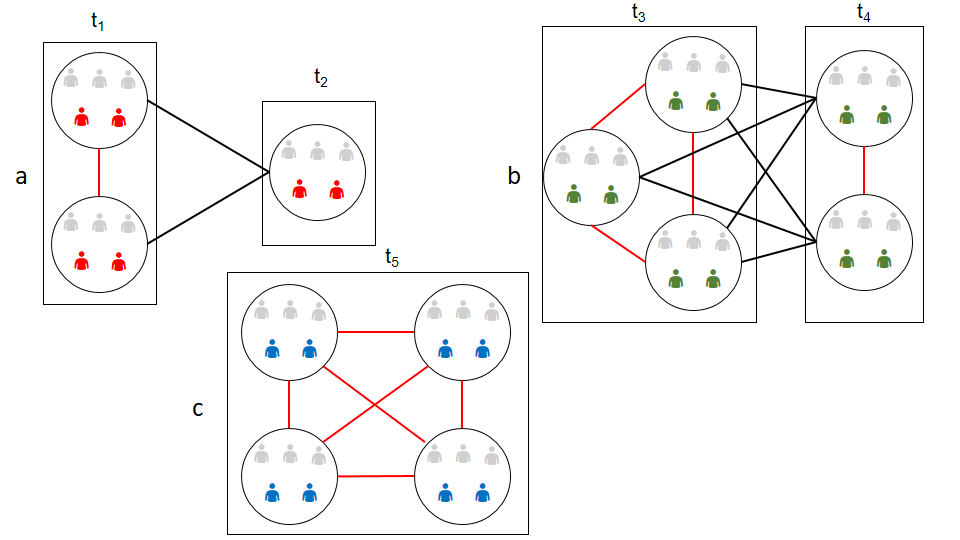}
	\caption{Examples of the connection strength between pairs of passengers in three different travel scenarios.. Rectangles represent vehicle trips and circles represent cliques. Edges marked with black increase the connection strength between highlighted passengers. Red edges penalize connection strength because these are on the same vehicle trip.  Figure 5/a: two passengers travel together in two cliques on vehicle trip $t_1$ and one clique on vehicle trip $t_2$, therefore $ g_{uv}^{t_{1}} = 2, \ g_{uv}^{t_{2}} = 1, \ g_{uv}=3$ and $s=2$. Figure 5/b: two passengers travel together in three cliques on $t_3$ and two cliques on $t_4$ making $g_{uv}^{t_{3}} = 3, \ g_{uv}^{t_{4}} = 2, \ g_{uv}=5$ and $s=6$. Figure 5/c: two passengers travel together on a single vehicle trip in four cliques making $ g_{uv}^{t_{1}} = 4, \ g_{uv}=4$ and $s=0$.}
\end{figure}

Algorithm 1 shows the construction of the {\em community network}, while Figure 5 illustrates a few examples for computing $s$ between passengers.
On Figure 5/a two passengers travel together in two cliques on vehicle trip $t_1$ and one clique on vehicle trip $t_2$, therefore $ g_{uv}^{t_{1}} = 2, \ g_{uv}^{t_{2}} = 1, \ g_{uv}=3$ and $s=2$. A different situation is shown on 5/b where two passengers travel together in three cliques on $t_3$ and two cliques on $t_4$ making $g_{uv}^{t_{3}} = 3, \ g_{uv}^{t_{4}} = 2, \ g_{uv}=5$ and $s=6$. Figure 5/c presents a trivial case, when two passengers travel together on a single vehicle trip in four cliques making $ g_{uv}^{t_{1}} = 4, \ g_{uv}=4$ and $s=0$.


\subsection{Passenger communities}

In this section we illustrate examples of passenger communities in the public transportation system of Twin Cities MN. The first example seen on Figure 6 shows subgraphs of the community network constructed from $G_{30}$, \textit{i.e.} the contact network only containing edges where contact duration is above 30 minutes. Figure 6/a depicts the entire community network. Most of the communities on this network are of size two or three, but there are several larger communities with strong connections between the members. Figure 6/b shows a subgraph where edges with weights $s<5$ are omitted as well as all nodes with degrees below two. The remaining subgraph contains the largest group of the network, while the largest individual community is depicted on 6/c.      

\begin{figure}[!htbp]
	\centering
	\includegraphics[width=0.9\textwidth]{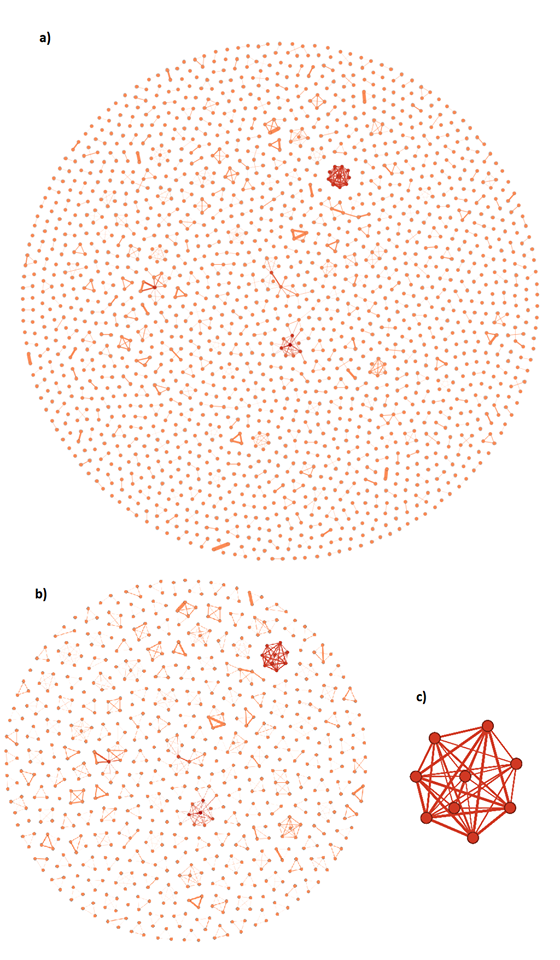}
	\caption{The community network of Twin Cities, MN. a) the whole community network, b) a subgraph with edge weights greater than 5, c) the largest passenger group of the network.}
\end{figure}

Figure 6/c shows the largest group in the network. The group contains nine passengers who traveled together on two different vehicle trips, while the overall time they spent using the public transportation network was almost 1.5 hours. The passengers embarked on route 805 in the morning between 7:08 and 7:16 near Blaine and traveled together to Northtown. They disembarked at 7:48 and waited together for the second vehicle 852 arriving at 8:12 and traveled together to downtown Minneapolis for almost an hour until 9:12 and 9:16. The travel path of the community, shown on Figure 7, indicates a commuting pattern from one of the suburbs to the city center of Minneapolis.

\begin{figure}[h]
	\centering
	\includegraphics[width=1.0\textwidth]{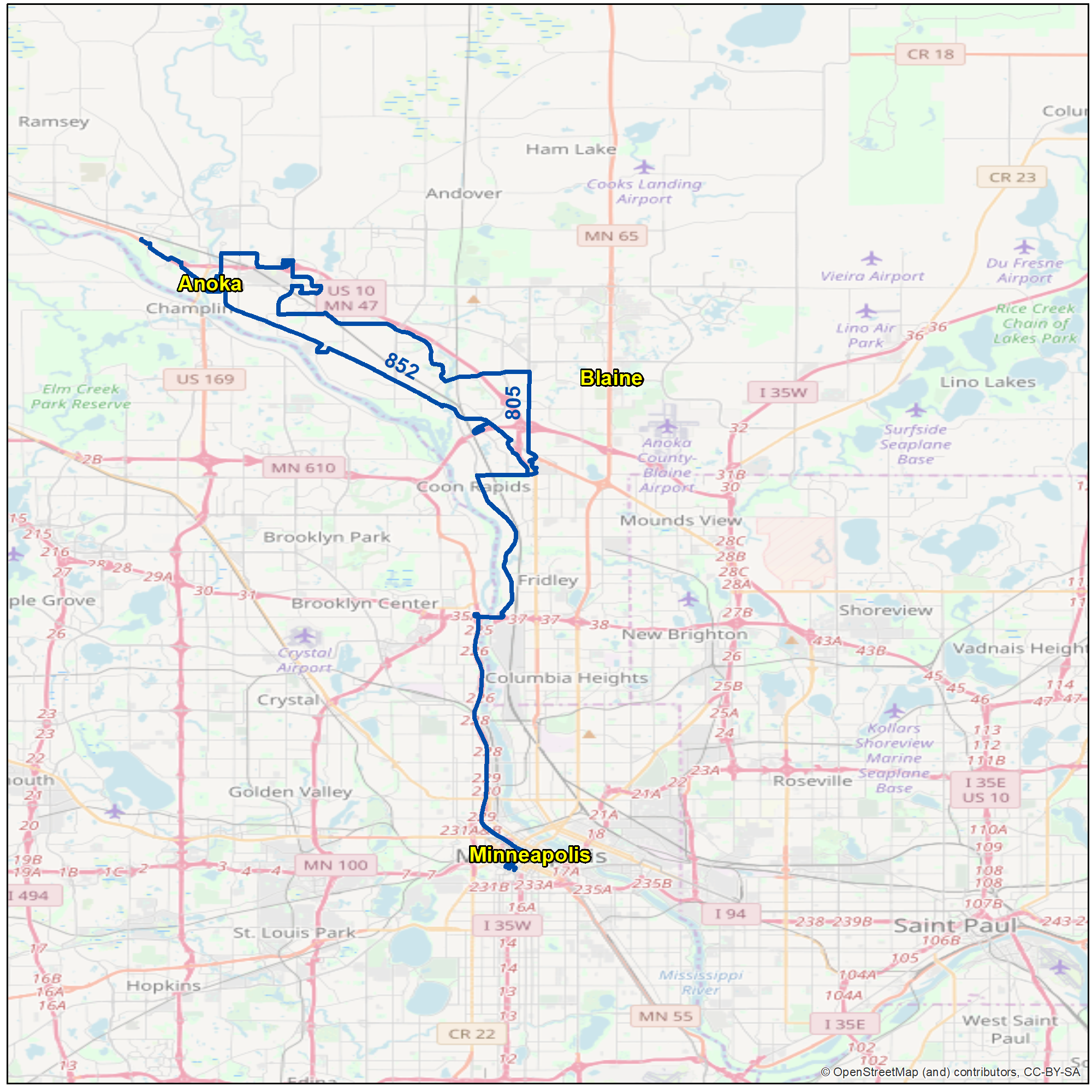}
	\caption{The travel path of the passenger community on Figure 6/c traveling from a suburb to the city center.}
\end{figure}

Both contact and community networks reveal contact patterns between passengers of a public transportation system. In the contact network, the weight between individual passengers is defined based on the contact duration on individual vehicle trips. In contrast, the community network provides a more refined way to represent connection strength which takes into account the amount of transfers passengers take together.

The underlying concept behind community structure in networks is homophily, which is a well-studied concept of the social sciences \cite{eagle2009inferring,yuan2006homophily,chin2012using}. Homophily states that people tend to form groups according to their lines of interest, occupation, etc. Physical proximity -- on public transportation for example -- is one of these indicators. Therefore, strong connections on the community network can help us uncover connections in other areas of life like workplace or school or other common interests. It should be noted, that physical proximity does not guarantee another type of connection, it simply increases the likelihood of occurrence for it.

\subsection{Epidemic spreading risk application}

One application of identifying the communities within the transit network, as described in this work, is infrastructure security. Understanding passenger communities enables more efficient and accurate tracking of infectious disease spread, were one to be naturally or maliciously introduced into the public transit system. 
One of the challenges in modeling epidemic spreading is accurately mapping the relationships between individuals traveling on the same vehicle.
A traditional contact network as defined in section 2 as well as in \cite{nets,sun2,trb} simply revelas the set of passengers who were present on the same vehicle and the amount of time they spend on the same vehicle. As shown in \cite{trb}, this limits the options for how infection spreading probabilities can be defined to be simply a function of contact duration, without the possibility to take into account physical proximity, communication etc. between people.

In contrast, the weights of the community network reveals a deeper level of connections between travelers, as passengers with strong connections in this network may also be connected in other areas of life. Passenger groups identified in the community network are more likely to be traveling within close physical proximity of each other and interacting with each other \cite{eagle2009inferring,yuan2006homophily,chin2012using}. As a consequence, the probablity of disease transmission between the travelers belonging to the same community is greater than between two passengers simply sharing the same vehicle without any other connection. This is especially true for large public transportation vehicles like trains or trams, where simply being present on the same vehicle may not imply any kind of connection at all. 

In this section we expand upon our previous work in \cite{nets} which examined epidemic spreading risk in the same public transportation network (Twin Cities, MN) with the goal of identifying the vehicle trips most likely to carry infected passengers. The analysis was presented in two parts. First, the passenger contact network, in the same form as in Section 2, was used to model a variety of outbreak scenarios. The scenarios differed in the number and distribution of initially infected passengers and the level of infectiousness, represented as the risk of spreading between passengers. The spreading risk was defined as the contact duration multiplied by a constant value. The scenarios were compared in terms of their impact on the network and confirmed previous observations in the literature, that is that the most central vehicle trips in the public transportation network are also the most susceptible to infection. The second part of the analysis focused on a newly proposed vehicle trip network, which represents the public transit network as a network of vehicle trips instead of passengers. We showed that centrality metrics on the vehicle trip network provided a more efficient way to detect the set of vehicle trips most susceptible to disease, and this estimation can be done much faster than running simulations on the contact network.

In the rest of this section we present an alternative way to model the risk of disease spreading between passengers, which exploits the community network structure introduced in this work. We define the infection transmission probability between a pair  of passengers to be a function of both the connection strength value used in the community network and contact duration. We believe this multidimensional transmission probability more accurately represents the level of connectivity between pairs of passengers. Using these new transmission functions, we implement similar spreading scenarios to the ones in \cite{nets} and identify and rank the vehicle trips susceptible to disease spreading. Below we define the new transmission probability function and the infection model, then present the new vehicle-trips identified to be at highest risk.

\subsubsection{Experiment setup}

In order to simulate an epidemic outbreak on the contact network we use the well-known discrete compartmental susceptible-infected (SI) model. In the SI infection process all nodes adopt one of two available states: susceptible (S) or infected (I). Real values denoted as edge infection probabilities are assigned to the links of the network and denoted as $w_e \in [0,1]$. The infection spreads in a network when susceptible nodes adopt an infected state. This is done in discrete time steps in an iterative manner starting from an initially infected set of nodes.  In each iteration each infected node tries to infect its susceptible neighbors according to the transmission probability of the link connecting them. If the attempt is successful, the susceptible node is transformed into an infected one in the next iteration. In this work we limit the number of iterations to five, representing a complete work week with recurrent commuting patterns. The inputs of this model are: 1. a contact network of individuals, 2. an assignment of weights to the links of the network which represent the infection transmission probabilities and 3. the set of initially infected nodes, e.g., individuals.

We use the original structure of the contact network, which connects all pairs of passengers that travel on a vehicle-trip together as the basis of this experiment. The link weights are computed in the following way. Since the nodes and links of the community network are a subset of the nodes and links of the contact network, if a link between two nodes is present in the community network we will assign its connection strength value to the corresponding link in the contact network. We do this in all cases where such an assignment can be made. In order to account for extreme outliers of connection strength we cap all values at 100, then rescale all values to the interval of $0.1$ and $0.8$ using a standard feature scaling method. The 0.8 upper bound is set because a transmission probability of 1.0 is assumed to be unrealistic. For all links of the contact network that do not have a corresponding link in the community network we assign a uniform infection value of $0.05$. In contrast to the duration-based value used in \cite{nets}, this enables us to capture the increased spreading risk between passengers sharing similar travel patterns, while still allowing disease to spread between travelers simply sharing a vehicle. Future work will explore the model's sensitivity to the uniform infection assignment using in this work. 

Following the procedure in \cite{nets}, we randomly select 100 passengers from the network to be initially infected. Due to the probabilistic nature of the simulation model, we run the SI infection model $k = 10000$ times to quantify the likelihood of each nod being in an infectious state at the end of the fifth time step.

\subsubsection{Vehicle trip ranking}

The infection model constructed above provides the likelihood of infection for each passenger in the contact network at the end of the simulation, i.e., after five days. As in \cite{nets}, we compute a similar infection value for vehicle trips by summing the probability of infection for all passengers on a given vehicle trip. While this does not represent a probability value in a strict sense, this value is proportional to the risk of getting infected on a given vehicle trip.

The routes which contain the highest risk vehicle trips, and corresponding travel times identified in this study are presented in Table 4. Figure 8 shows the routes on the map of the city. Coinciding with the findings in Section 3 and also in \cite{nets}, the vehicle trips are in the morning and afternoon peak hours. Also coinciding with the results in \cite{nets} all of the high risk routes identified go through the city center. 

Since the goal of Section 3.2 -- identifying the most frequent trip combinations -- and the goal of this section -- identifying the most risky vehicle trips in the case of an epidemic outbreak -- are different, the difference between the selected vehicle trips are not surprising. We observe some similarities between the "highest risk" vehicle trips identified here, and the most frequent vehicle trip combinations (identified in section 3.2). For example, route 18 connects Bloomington with the city center, while route 61 connects St. Paul to the same destination. We have seen in Section 3.2, that these target-destination pairs are present in the most frequent combinations as well. As these routes connect the most important parts of the city, some similarity is expected.

\begin{table}[!h]
	\begin{center}
		\caption{Ranking of vehicle trip where infection is most likely to appear.}\label{tab:test}
		\begin{tabular}{r|c c} \hline
			Rank & Route number & Start time\\ \hline
			$1.$ & 25   & 6:52 \\ 
			$2.$ & 17   & 6:13  \\
			$3.$ & 25   & 5:11  \\
			$4.$ & 14  & 16:54 \\
			$5.$ & 5  & 5:36 \\ 
			$6.$ & 25  & 7:16 \\
			$7.$ & 18  & 6:07 \\
			$8.$ & 11  & 15:52 \\
			$9.$ & 11 & 17:14 \\
			$10.$ & 61  & 6:07 \\ \hline
		\end{tabular}
	\end{center}
\end{table}

The set of high risk vehicle trips identified here partially correspond to those identified in \cite{nets}. Specifically, the methodology proposed in \cite{nets} identifies vehicle trips on the routes 7, 9, 11, 14, 17, 25, 61 and 94 as most at risk. These are routes crossing or connecting to the city center in the peak hours, so even though they do not completely match those in Table 4, the pattern they present is the same. 
The similarity in the set and type of routes identified in both studies points to the critical role of the network structure in modeling outbreak risk. Future research will continue to build on this application, and further explore the robustness and sensitivity of the proposed methodology. The relevant sensitivity analysis is, however, outside the scope of this work.

\begin{figure}[h]
	\centering
	\includegraphics[width=1.0\textwidth]{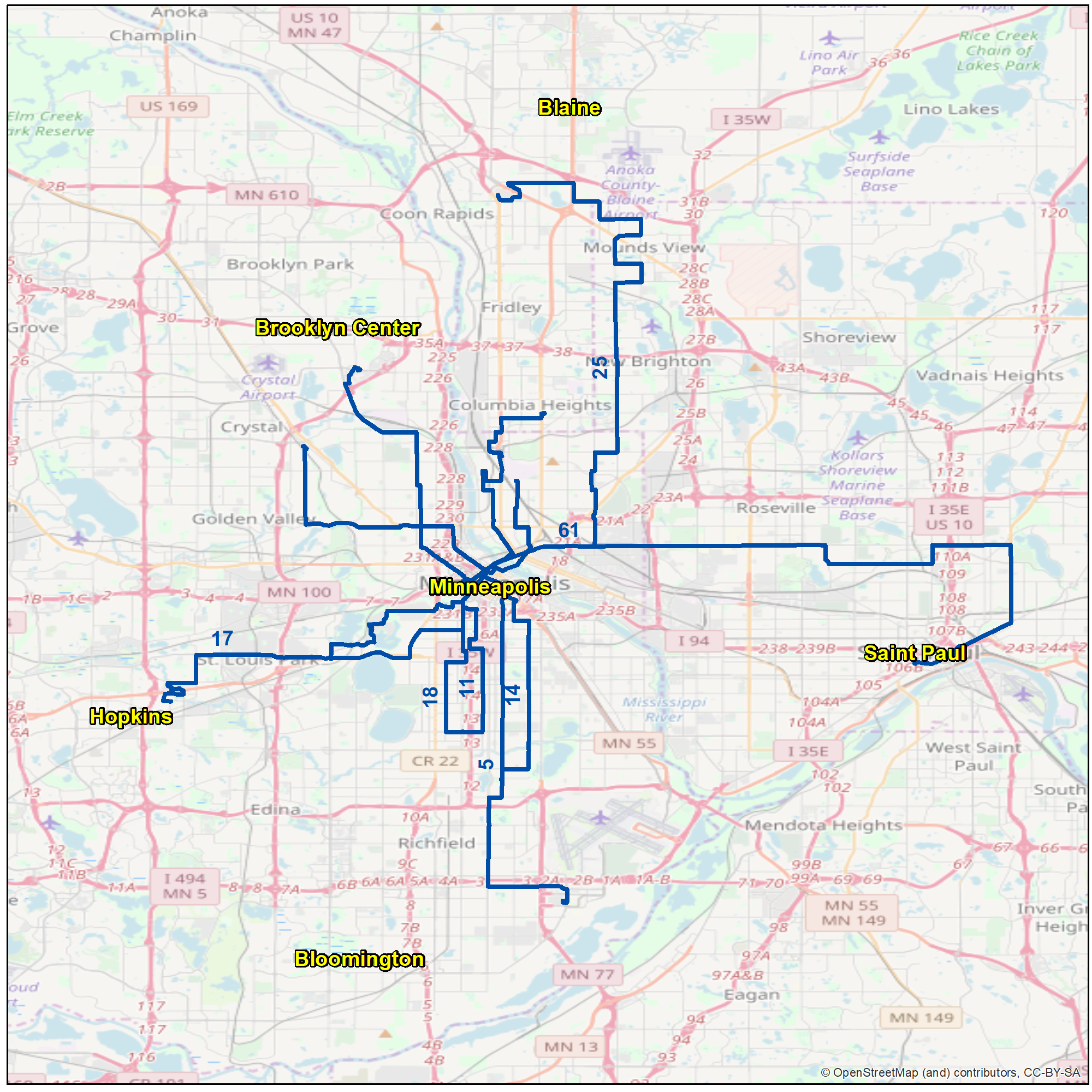}
	\caption{Vehicle trips most likely to carry infected passengers in the public transit system of Twin Cities, MN.}
\end{figure}



\section{Conclusions}

In this paper we proposed two novel network structures to detect and quantify the movement patterns of passengers of a public transit system. The transfer network tracks the movements of atomic passenger group while the community network links passengers with similar travel patterns. We presented applications for both networks, specifically to identify the most frequently used travel paths, i.e., routes and transfers, and model epidemic risk posed by passengers of a public transit network, respectively.

The transfer network was demonstrated as a tool to efficiently detect the most frequently used vehicle trip combinations in the public transportation system. We have shown that the most frequent combinations follow a similar pattern. The trip pairs identified using the transfer network identified peak morning and afternoon trips that connect the outlying suburbs with the CBD, which corresponds to existing literature.

We further illustrated how the community network can be used to augment the traditional contact network structure, and reveal components of the transit system at risk from an epidemic outbreak.  Similar to \cite{nets} we sought to identify the vehicle trips most likely to carry infected passengers. However, in contrast to the model presented in \cite{nets} where infection probability between passengers was solely based on contact duration, here we used the connection strength value to estimate physical contact between passengers in an alternative way. 
Our results reinforce previous observations, that vehicle trips crossing or connecting to the city center pose the highest risk.

This study is subject to certain limitations. We note, that the transit demand model used as an input of both works was used to generate commuting patterns for single workday. This limitation is most critical for the epidemic application due to the implicit assumption that daily travel patterns remain constant during a five day work week. While this assumption is supported by a recent study \cite{sun2}, more long term observations potentially involving weekends and public holidays would improve the quality of the results presented in this paper.

Further, while we present an alternative metric to quantify disease transmission risk between passengers (compared to contact duration alone), without any real-world observations on an outbreak validating the results of this work remains a challenge, and will be the focus of future research. 

\begin{acknowledgements}
The authors are grateful to Metropolitan Council of Twin Cities for sharing the activity-based travel demand model with researchers at the University of Minnesota. Any limitations of this study remains the sole responsibility of the authors.
L\'aszl\'o Hajdu acknowledges the support of the National Research,
Development and Innovation Office - NKFIH Fund No. SNN-117879.
Mikl\'os Kr\'esz acknowledges the European Commission for funding the
InnoRenew CoE project (Grant Agreement \#739574) under the Horizon2020
Widespread-Teaming program and the support of the EU-funded Hungarian grant
EFOP-3.6.2-16-2017-00015
\end{acknowledgements}

\bibliographystyle{spmpsci}      



\end{document}